\documentclass[a4paper,11pt]{article}

\pdfoutput=1

\newcommand{\mtvAuthors}{Bosco, C., de Rigo, D., Dewitte, O., Montanarella, L.}
\newcommand{\mtvYear}{2011}
\newcommand{\mtvTitle}{Towards the reproducibility in soil erosion modeling: a new Pan-European soil erosion map.}
\newcommand{\mtvTITLE}{Towards the reproducibility in soil erosion modeling:\\a new Pan-European soil erosion map}
\newcommand{\mtvJournal}{{\em Wageningen Conference on Applied Soil Science ``Soil Science in a Changing World''}, 18~-~22~September~2011, Wageningen, The Netherlands.}

\newcommand{\mtvAbstract}{%
This is the authors' version of the work. It is based on a poster presented at the Wageningen Conference on Applied Soil Science, \\
{\url{http://www.wageningensoilmeeting.wur.nl/UK/}}\\[12pt]%
Cite as:%
\\[7pt]%
{\sloppy \parbox[l]{138mm}{
\mtvAuthors{}, \mtvYear{}. 
   {\bf \mtvTitle{}} 
   {\em Wageningen Conference on Applied Soil Science ``Soil Science in a Changing World''}, 18 - 22 September 2011, Wageningen, The Netherlands.}
}\\[19pt]
Author's version DOI: \href{http://dx.doi.org/10.6084/m9.figshare.936872}{10.6084/m9.figshare.936872}%, arXiv: \href{http://arxiv.org/abs/1311.4762}{1311.4762}
\\[12pt]
\mtvBf{Abstract}\\
Soil erosion by water is a widespread phenomenon throughout Europe and has the potentiality, with his on-site and off-site effects, to affect water quality, food security and floods. Despite the implementation of numerous and different models for estimating soil erosion by water in Europe, there is still a lack of harmonization of assessment methodologies. \\[1mm]
Often, different approaches result in soil erosion rates significantly different. Even when the same model is applied to the same region the results may differ. This can be due to the way the model is implemented (i.e. with the selection of different algorithms when available) and/or to the use of datasets having different resolution or accuracy. Scientific computation is emerging as one of the central topic of the scientific method, for overcoming these problems there is thus the necessity to develop reproducible computational method where codes and data are available. \\[1mm]
The present study illustrates this approach. Using only public available datasets, we applied the Revised Universal Soil loss Equation (RUSLE) to locate the most sensitive areas to soil erosion by water in Europe. \\[1mm]
A significant effort was made for selecting the better simplified equations to be used when a strict application of the RUSLE model is not possible. In particular for the computation of the Rainfall Erosivity factor (R) the reproducible research paradigm was applied. The calculation of the R factor was implemented using public datasets and the GNU R language. An easily reproducible validation procedure based on measured precipitation time series was applied using MATLAB language. Designing the computational modelling architecture with the aim to ease as much as possible the future reuse of the model in analysing climate change scenarios is also a challenging goal of the research.
}

\usepackage{mtvstyle}
\newcommand{\mtvSection}[1]{\section*{#1}}
\newcommand{\mtvSubsection}[1]{\subsection*{#1}}
\newcommand{\mtvBf}[1]{\textsf{\textbf{#1}}}

\author[1]{{\Large \textsf{Claudio Bosco}}~}
\author[1,2]{{\Large \textsf{Daniele de Rigo}}~}
\author[1]{{\Large \textsf{Olivier Dewitte}}~}
\author[1]{{\Large \textsf{Luca~Montanarella}}~ \vspace{2mm}}

\affil[1]{\small \;European Commission, Joint Research Centre, Institute for Environment and Sustainability\\

Via E. Fermi 2749, I-21027 Ispra (VA), Italy\smallskip }
\affil[2]{\;Politecnico di Milano, Dipartimento di Elettronica e Informazione\\

Via Ponzio 34/5, I-20133 Milano, Italy\vspace{-4mm}}
\date{}

\begin{document}

  \maketitle

  \mtvMakeAbstract{}
  
\newpage\mbox{ }\vspace{2mm}

\begin{mtwocols}
\mtvSection{Introduction}

Despite the implementation of a variety of models for estimating soil erosion by water in Europe \cite{Rusco_etal_2008}, there is still a lack of harmonization of assessment methodologies.

\medskip

Often, distinct approaches lead to significantly different soil erosion rates and even when the same model is applied to the same region the results may differ. This can be due to the way the model is implemented (i.e. with the selection of different algorithms when available) and/or to the use of datasets having distinct resolution or \mbox{accuracy}. 

\medskip

Scientific computation is emerging as one of the central topic within environmental modelling \cite{Casagrandi_and_Guariso_2009}, to overcome these problems there is thus the necessity to develop reproducible computational methods based on free software and data \cite{Stallman_2005,Waldrop_2008}, and to also reuse -- in a controlled way -- empirical equations for compensating the lack of detailed~data. 

\medskip

The present study illustrates such an approach. Using only public available datasets (SGDBE \cite{Heineke_etal_1998}, SRTM \cite{Farr_etal_2007}, CLC and E-OBS \cite{Haylock_etal_2008}) , we applied a derived version of the Revised Universal Soil loss Equation (RUSLE) \cite{Renard_etal_1997} to locate the most sensitive areas to soil erosion in Europe. We decided to use a RUSLE-based approach because of the flexibility and least data demanding of the model \cite{Bosco_etal_2008,Bosco_etal_2009}.

\medskip

A significant effort was made \cite{de_Rigo_Bosco_2011,Bosco_etal_2011} toward reproducibility and to select the better simplified equations to be used when a strict application of the model is not possible. In particular for the computation of the Rainfall Erosivity factor (R) the reproducible research paradigm was applied.

\vspace{7mm}
\mtvSection{The model}

The Revised Universal Soil Loss Equation (RUSLE) has been extended by including a correction factor $St_{c,Y}$ able to consider the stoniness: 

\[
\begin{array}{lcl}
Er_{c,Y} & = & R_{c,Y} \:\cdot\: K_{c,Y} \:\cdot\: L_{c,Y} \:\cdot\: S_{c,Y} \:\cdot\: \\[1mm]          
&& C_{c,Y} \:\cdot\: St_{c,Y} \:\cdot\: P_{c,Y}
\end{array}
\]  

\smallskip

\noindent where the factors refer to a specific grid cell $c$ and represent the annual average for a certain set of years $Y = {y_1, \cdots ,y_i,\cdots, y_{n_Y}}$ (R factor) or -- where data are stable or missing -- the values corresponding to a temporally more localized set of data:

\newcommand{\mytabspA}{1.5mm}
\begin{equation*}
\begin{array}{lcl}
Er_{c,Y}    &=& \text{average annual soil loss } \\&&(t\: ha^{-1} \:yr^{-1}). \\[\mytabspA]
R_{c,Y}     &=& \text{rainfall erosivity factor } \\&&(MJ\: mm\: ha^{-1}\: h^{-1}\: yr^{-1}). \\[\mytabspA]
K_{c,Y}     &=& \text{soil erodibility factor } \\&&(t\: ha\: h\: ha^{-1}\: MJ^{-1}\: mm^{-1}). \\[\mytabspA]
L_{c,Y} 	&=& \text{slope length factor} \\&&\text{(dimensionless).} \\[\mytabspA]
S_{c,Y}	    &=& \text{slope steepness factor} \\&&\text{(dimensionless).} \\[\mytabspA]
C_{c,Y}	    &=& \text{cover management factor} \\&&\text{(dimensionless).}\\[\mytabspA]
St_{c,Y} 	&=& \text{stoniness correction factor} \\&&\text{(dimensionless).}\\[\mytabspA]
P_{c,Y}     &=& \text{support practice aimed at} \\&&\text{erosion control (dimensionless).}\\[2mm]
\end{array}
\end{equation*}

\medskip

\noindent\mtvBf{Advantages}: simplicity and robustness.

\medskip

\noindent\mtvBf{Limits}: at this resolution and according to the uncertainties associated with the input data, this model is only relevant to locate the areas prone to soil erosion.

\vspace{5mm}\mbox{ }
\end{mtwocols}

\vspace{7mm}\mbox{ }

\newcommand{\mytabspB}{2mm}
\begin{table}
\mtvBf{\caption{Public available datasets used for running the extended RUSLE model} }
\vspace{4mm}
\centering
\begin{tabular}{p{20mm}p{45mm}p{55mm}}
\mtvBf{Factor} & \mtvBf{Data} & \mtvBf{Database} \\[\mytabspB]
R \cite{Renard_etal_1997,Bollinne_etal_1979,Ferro_etal_1999,Loureiro_and_Coutinho_2001,Rogler_Schwertmann_1981}
& Average daily precipitation
& The European daily gridded \phantom{\hspace{5mm}} dataset -- E-OBS \\[\mytabspB]
K \cite{Renard_etal_1997}  
& Topsoil silt, clay, sand \% 
& The database of European soils -- SGDBE\\[\mytabspB]
L \cite{Nearing_1997}
& Elevation
& SRTM 90 m \\[\mytabspB]
S \cite{Nearing_1997}
& Elevation
& SRTM 90 m \\[\mytabspB]
C \cite{Morgan_2005,Suri_2002,Cebecauer_and_Hofierka_2008}
& Land cover classes
& CORINE Land Cover\\[\mytabspB]
St \cite{Poesen_etal_1994} 
& Percentage of stoniness 
& The database of European soils -- SGDBE\\[\mytabspB]
P
& Set equal to 1
& --- \\[-5mm]
\end{tabular}
\end{table}

\begin{mtwocols}
\mtvSubsection{The implemented reproducible part of the model}

\noindent
\mtvBf{Rainfall erosivity factor}. One of the main factors influencing soil erosion by water is the rainfall intensity. The $R$ factor measures the erosivity of precipitations. 
The composite parameter $EI^{30}$ has been identified by Wischmeier \cite{Wischmeier_1959} as the best indicator of precipitation erosivity.
For determining $EI^{30}$ the kinetic energy $E$ of rain is multiplied by the maximum rainfall intensity $I^{30}$ occurred in 30 minutes in every $k$-th precipitation event of the $i$-th year.

\medskip
The R factor represents the average, on a consistent set of data, of $n_Y$ sums of $EI^{30}$ values. Each sum is computed for the whole set of $n_{y_i}^{\text{event}}$ precipitation events in the $i$-th year:
 
\[
\begin{array}{ll}
R_{c,Y} 
\quad &= \quad
\displaystyle \frac{1}{n_Y} \cdot \sum_{i=1}^{n_Y} \sum_{k_i = 1}^{n_{y_i}^{\text{event}}} E_{c,k_i} \cdot I_{c,k_i}^{30} \\[4mm]
\quad &= \quad 
\displaystyle \frac{1}{n_Y} \cdot \sum_{i=1}^{n_Y} \sum_{k_i = 1}^{n_{y_i}^{\text{event}}} EI_{c,k_i}^{30}
\end{array}
\]

\noindent
Within the framework, the complete equation has been fully implemented to accurately estimate R where detailed time series of measured precipitation (10 to 15 minutes of time-step) have been made available across Europe.

\medskip
However, the scarcity of these accurate datasets and the desire to design a reusable framework for assessing water soil erosion at regional scale with only limited and approximated information motivated the creation of a climatic-based ensemble model for estimating erosivity from multiple available empirical relationships. 

\medskip
The array programming paradigm \cite{Iverson_1980,Quarteroni_Saleri_2006} was applied using MATLAB language \cite{MathWorks_2011} and GNU Octave \cite{Eaton_etal_2008} computational environment. Within that paradigm, a semantic-constraint oriented support was adopted by exploiting the Mastrave library \cite{de_Rigo_SemAP_Mastrave,de_Rigo_exp2012}.

\begin{figure*}[htp]\centering 
\includegraphics[width=14.5cm]{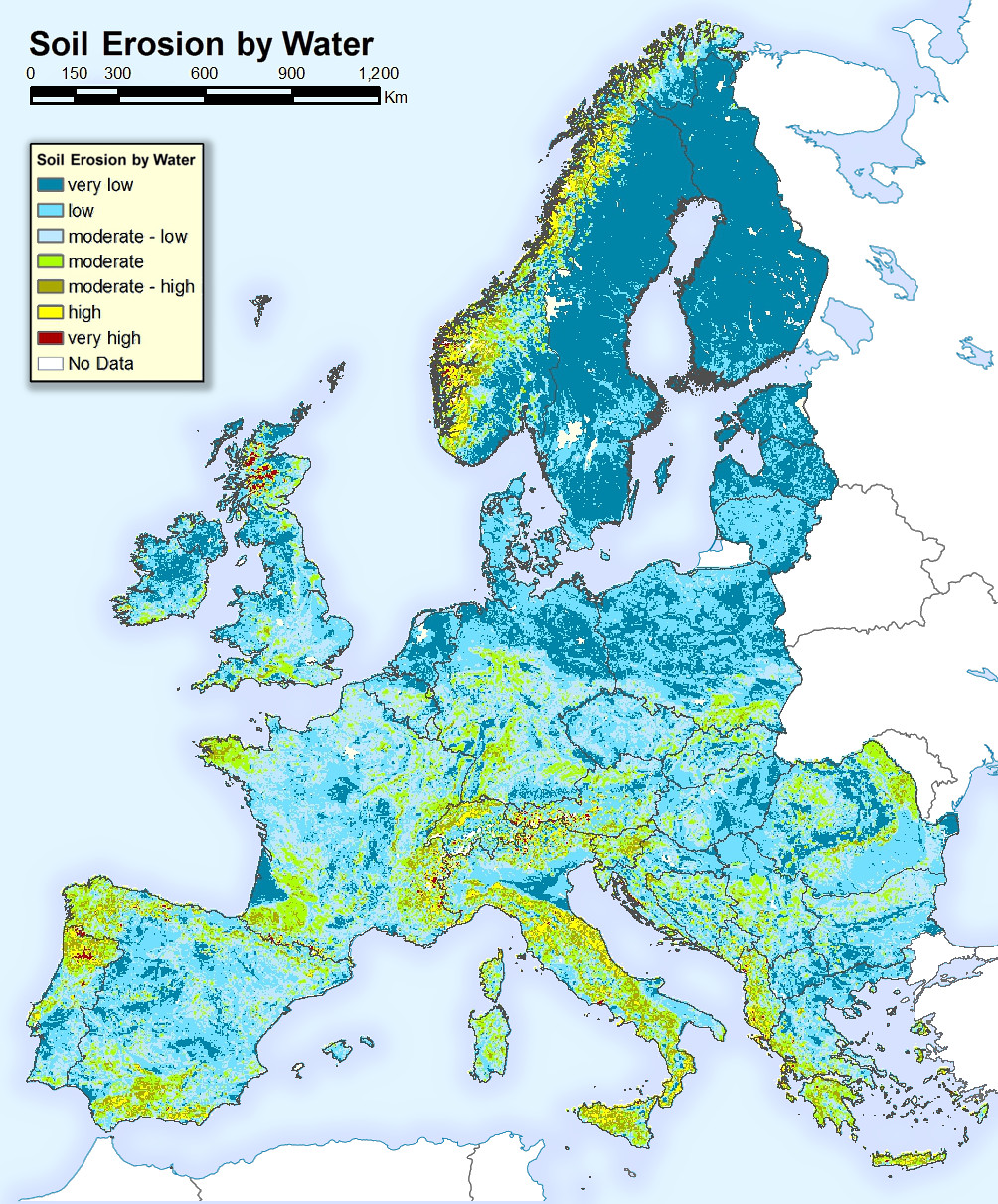}
\mtvBf{\caption{\small Soil erosion rate by water $(t ha^{-1} yr^{-1})$ estimated applying the extended RUSLE model.}}
\label{fig:fig1}
\end{figure*}

\begin{figure*}[ht]\centering 
\includegraphics[width=14.5cm]{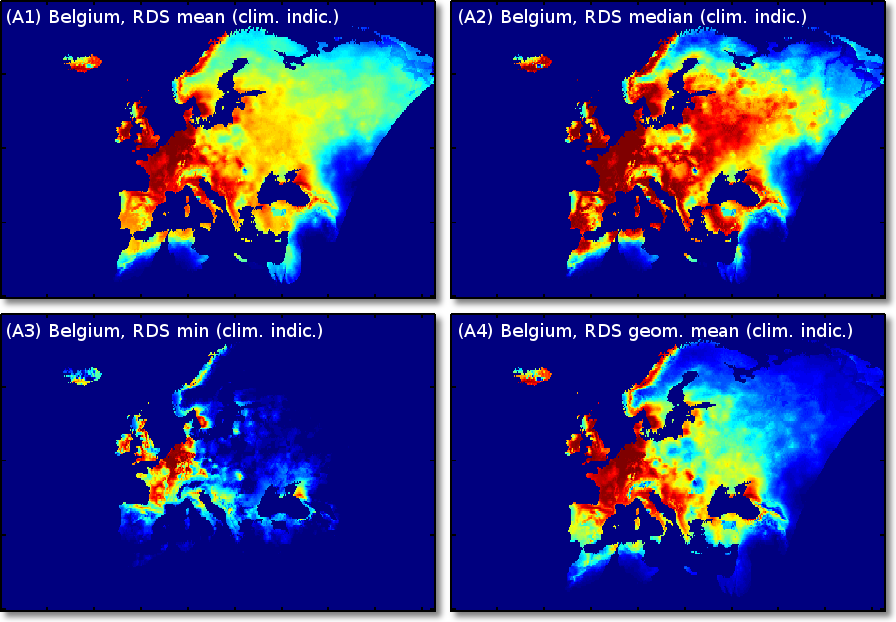}
\mtvBf{\caption{\small Climatic similarity estimated applying the Relative Distance Similarity (RDS) to the Bollinne equation (Belgium) for rainfall erosivity. The similarity of 26 climatic indicators over the whole Europe is shown (red: maximum similarity; blue: maximum dissimilarity) and aggregated computing respectively the mean (A1), median (A2), minimum (A3) and geometric mean (A4).}}
\label{fig:fig2}
\end{figure*}

\medskip
Multiple layers of geospatial data over a wide spatial extent may naturally be modelled as corresponding arrays (e.g. here raster grids of heterogeneous - coarser or denser -  spatial resolution have been used). Geoprocessing is required for the layers to be transformed in arrays with harmonised projection and datum. 

\medskip
Array programming has been introduced by Iverson \cite{Iverson_1980} in order for the gap between algorithm implementation and mathematical notation to be mitigated. As Iverson underlined, ``the advantages of executability and universality found in programming languages can be effectively combined, in a single coherent language, with the advantages offered by mathematical notation'' \cite{Iverson_1980}.

\medskip
Following this approach, prototyping  complex algorithms can benefit from a compact array-based mathematical semantics. This way, the mathematical reasoning is relocated directly into the source code, actually the only place where the mathematical description is completely formalised and reproducible.

\medskip
The semantic array programming paradigm \cite{de_Rigo_SemAP_Mastrave,de_Rigo_exp2012} (here applied \cite{Bosco_etal_inprep}) has been designed to support nontrivial scientific modelling with the help of two additional design concepts:

\begin{itemize}
\item {\em modularizing} complex data-transforma\-tions in autonomous tasks by means of general and concise sub-models, possibly suitable of reuse in other context. A harmonised predictable convention in module interfaces also relies on self-documenting the code;

\item {\em semantically constraining} the information flow in each module (input and output variables and parameters) instead of relying on external assumptions (e.g. instead of assuming the correctness of input information structured as an object).

\end{itemize}

\noindent In the present application, the R factor climatic-based ensemble model was implemented using public datasets and a novel methodology was applied for merging together multiple empirical equations. This was done by extending the original geographical domain of validity of each equation to similar areas.

\medskip The climatic similarity has been based on the relative-distance similarity methods of Mastrave \cite{de_Rigo_SemAP_Mastrave}. The climatic layers have been computed by using GNU R language \cite{R_DevelCoreTeam_2005} and GNU Octave. The R factor computational framework will be available as free software \cite{Stallman_2009}.

\mtvSubsection{Climatic ensemble modelling using Relative-Distance Similarity}

The ensemble modelling procedure was applied to 7 empirical equations based on significant correlations between climatic information (such as average annual precipitation, Fournier modified index, monthly rainfall for days with $\ge 10.0\, mm\,$, ...) and locally measured erosivity of 4 geographical areas: Algarve (Portugal), Belgium, Bavaria (Germany) and Sicily (Italy) \cite{Bollinne_etal_1979,Ferro_etal_1999,Loureiro_and_Coutinho_2001,Rogler_Schwertmann_1981}.  

\medskip
Similarity maps with respect to the climatic conditions of each equation's geographical domain have been computed based on the relative distance (dimensionless) between pan-European maps of 26 climatic indicators and the corresponding indicators' values of the equation area of validity. The behaviour of each empirical equation outside its definition domain was also assessed for preventing meaningless out-of-range values to degrade the ensemble estimation.

\medskip  The aggregated similarities for each equation have been normalized for estimating the ensemble erosivity map using weighted median \cite{de_Rigo_wmedian,de_Rigo_SemAP_Mastrave} of the 7 empirical models.  

\medskip The contribution of each empirical equation based on its aggregated similarity was accounted to estimate a qualitative trustability map of the ensemble generalization. As a whole, the ensemble model is therefore a reproducible, unsupervised data-transformation model applied to climatic data to reconstruct erosivity.

\begin{figure*}[ht]\centering 
\includegraphics[width=14.5cm]{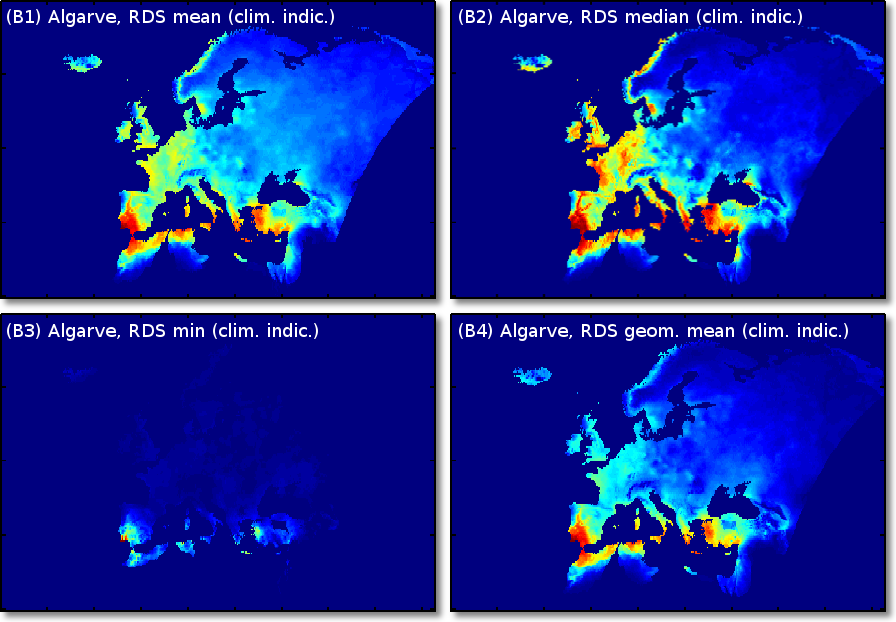}
\mtvBf{\caption{\small Climatic similarity estimated applying the Relative Distance Similarity (RDS) to the equation of de Santos Loureiro and de Azevedo Coutinho (Algarve) for rainfall erosivity. The similarity of 26 climatic indicators over the whole Europe is shown (red: maximum similarity; blue: maximum dissimilarity) and aggregated computing respectively the mean (B1), median (B2), minimum (B3) and geometric mean (B4).}}
\label{fig:fig3}
\end{figure*}

\mtvSection{Conclusions}

A lightweight architecture has been proposed to support environmental modelling within the paradigm of semantic array programming \cite{de_Rigo_SemAP_Mastrave,de_Rigo_exp2012}. The applied methodology benefits from the array programming paradigm with semantic constraints to concisely implement models as semantically enhanced composition of interoperable modules. 

An application for estimating the pan-European soil erosion by water, using a revised version of the RUSLE model, has been carried out merging existing empirical rainfall-erosivity equations within a climatic ensemble model based on the novel relative-distance similarity. An accurate estimation of the rainfall erosivity factor, applying the proposed architecture, has been implemented and will be used for validating simplified R-factor equations. 

\mtvSection{Next Steps}

The proposed architecture is designed to ease the future integration, within the same lightweight framework, of erosion-related natural resources models \cite{de_Rigo_Bosco_2011,Bosco_etal_inprep}. In particular, forest resources and wildfires\cite{Shakesby_2011}, natural vegetation \cite{Zuazo_Pleguezuelo_2009} and agriculture will be considered as key land cover factors under different climate change scenarios.

\medskip
\noindent\mtvBf{Acknowledgments}. We acknowledge the E-OBS dataset from the EU-FP6 project ENSEMBLES (\url{http://ensembles-eu.metoffice.com}) and the data providers in the ECA\&D project (\url{http://eca.knmi.nl}).
\end{mtwocols}

\medskip
\checkendpaper{}

\begin{footnotesize}

\raggedright
\nohyphens{

}
\end{footnotesize}

\end{document}